\documentclass[overcite,twocolumn,showpacs,aps,prl,amsmath]{revtex4}
\usepackage{epsfig}

\newcommand\kessence{\textit{k}-essence}
\newcommand\CMB{\textsc{cmb}}
\newcommand\CMBFAST{\textsc{cmbfast}}
\newcommand\CDM{\textsc{cdm}}

\begin{document}

\title{Effects of the Sound Speed of Quintessence on 
the Microwave Background and Large Scale Structure}
\author{Simon DeDeo$^1$, R.R.~Caldwell$^2$, Paul~J.~Steinhardt$^3$}
\date{\today}

\pacs{PACS number(s): 98.80.-k, 98.70.Vc, 98.65.-r}

\affiliation{$^1$Department of Astrophysical Sciences, Princeton University, Princeton, NJ 08544, USA \\
$^2$Department of Physics \& Astronomy, Dartmouth College, Hanover, 
NH 03755, USA \\
$^3$Department of Physics, Princeton University, Princeton, NJ 08544, USA }

\begin{abstract}

We consider how quintessence models in which the sound speed
differs from the speed of light  and varies with time 
affect the
cosmic microwave background and the fluctuation power
spectrum. Significant
modifications occur on length scales related to the
Hubble radius  during epochs in which the sound speed is
near zero
and the quintessence contributes a
non-negligible fraction of the total energy density. 
For the microwave background, we find that the usual 
enhancement of the lowest multipole moments by the integrated
Sachs-Wolfe effect can be modified, resulting in suppression
or bumps instead. Also, the sound speed can produce
oscillations and other effects  at
wavenumbers $k > 10^{-2}$~h/Mpc in the fluctuation power
spectrum.

\end{abstract}

\maketitle


One of the greatest challenges in cosmology today is to identify the nature of
the dark energy component that comprises most of the energy density of the
universe and that is causing the expansion of the universe to
accelerate.\cite{Bahcall} Two candidates are a cosmological constant (or vacuum
density) and quintessence,\cite{Cal98} a dynamical energy component with
negative pressure. Distinguishing the two is important for cosmology in order
to refine our knowledge of the composition of the universe and to trace more
accurately its evolution. It is even more important for fundamental physics
since it informs us how we must modify unified theories to incorporate dark
energy.

One way to distinguish whether the dark energy is due to  a cosmological
constant or quintessence is to measure  the equation of state, $w$, the ratio
of the pressure $p$ to the energy density $\rho$. A cosmological constant
always has $w=-1$ whereas a scalar field generally has a $w(z)$ that  differs
from unity and varies with red shift $z$. Through measurements of supernovae,
large-scale structure and the cosmic microwave background (\textsc{cmb})
anisotropy, the equation of state may be determined accurately enough in the
next few years to find out whether $w$ is different from $-1$ or not.

A second way to distinguish the nature of dark energy is to measure its sound
speed to determine if it is different from unity. The sound speed can be
detected because it affects the perturbations in the quintessence energy
distribution. This approach is less generic because the sound speed in many
models of quintessence in the literature is equal to unity (the speed of
light),  {\it e.g.}, models in which  quintessence  consists of a scalar field
($\phi$) with canonical kinetic energy density ($X\equiv \frac{1}{2}(\partial
_{\mu }\phi )^{2}$) and a positive potential energy density ($V(\phi)$).
However, in general, the sound speed can differ from unity and  vary with time.
Detecting these effects is an independent  way of showing that dark energy does
not consist of a cosmological constant.

An important motivating example is \textit{k}-essence.\cite{kess} In these
models, the \textit{k}-essence\ undergoes two transitions in its behavior, one
beginning at the onset  of matter-domination and a second when
\textit{k}-essence\ overtakes the matter density. During the
radiation-dominated era, the  $k$-essence energy tracks the radiation, falling
as $1/a^4$ where $a$ is the scale factor. This tracking feature is significant
because it explains how dark energy can remain a subdominant but not completely
negligible component prior to matter-radiation equality. Generic models with
this behavior ($k$-essence is only one example) are known as
``trackers."\cite{tracker} The tracking property is relevant to our purpose
here because the sound speed can have greater influence if the dark energy is
non-negligible for a broad range of red shift. (A cosmological constant
becomes  completely negligible by red shift $z=2$, but this is not the case for
typical tracker-type models.)

What is special about \textit{k}-essence\ models is that the onset of the
matter-dominated era  automatically triggers a change in the  behavior of
\textit{k}-essence\ such that it begins to act as an energy component with
$w(z)  \approx -1$. The effect is achieved with a scalar field by introducing
a  kinetic energy density which  contains higher order derivative terms (that
is, terms non-linear in $X$) and  which results in attractor solutions with the
desired behavior.   This kind of model is appealing because it explains
dynamically why cosmic acceleration only begins well after matter-domination,
just as we observe, which is one of  the puzzling aspects of  dark energy. When
\textit{k}-essence\ overtakes the matter density, $w(z)$ changes to another
value greater than $-1$, the precise value of which depends on the detailed
model. The way in which the attractor behavior of $k$-essence is automatically
controlled by the  background equation of state is  important  motivation for
cosmologists to  consider  $k$-essence models,  specifically.  However,  the
analysis in this paper is more general.

\begin{figure}
\begin{center}
\epsfig{file=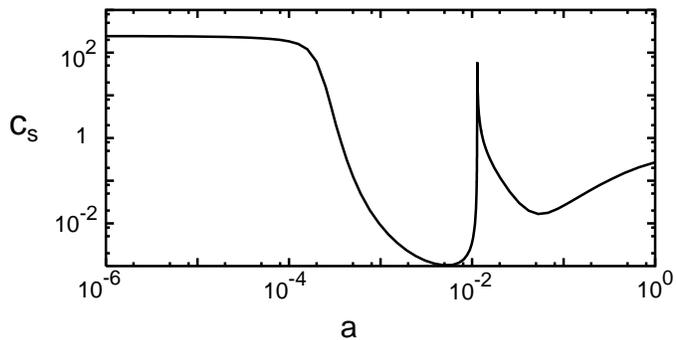,width=3.5in}
\end{center}
\caption{The speed of sound as a function of the scale factor $a$
($a_{\text{today}}=1$) for a \textit{k}-essence model studied in
Ref.\protect{\cite{kess}}. Note the changes in  $c_s$ between the last
scattering surface and  the present epoch triggered by the transformations in
the background equation of state.}
\label{fig:fig1}
\end{figure}

 For the purposes of this paper, what is significant about the $k$-essence
 example is an indirect consequence of the tracker behaviour: the sound speed
 $c_s$ undergoes dramatic changes as $w(z)$ passes through a series of
 transitions in evolving from the radiation- to the matter- to the
 $k$-essence-dominationed epochs. The behaviour of $c_s$ in a particular
 $k$-essence model\cite{kess} is shown in Fig.~\ref{fig:fig1}. Here the sound
 speed (\emph{i}) is relativistic in the radiation-dominated epoch, (\emph{ii})
 has rapid changes between zero and relativistic as $w(z)$ settles towards the
 attractor solution, which corresponds to $c_S^2 \rightarrow 0$ and $w(z)
 \rightarrow -1$, and, then, (\emph{iii}) increases somewhat at recent red
 shift when $k$-essence overtakes the matter density. The equation of state in
 tracker models must always have these three phases. Without knowing what the
 dark energy consists of, it is reasonable to imagine a generic class of dark
 energy models where the sound speed during each phase varies in a manner
 similar to the specific model described here. In particular, changes in sound
 speed are related to the changes in the scalar field as it moves rapidly
 between the radiation tracking and attractor solution in phase (\emph{ii}),
 and asymptotes to the $w=-1$ attractor point (\emph{iii}).

Our study here does not rely specifically on non-linear kinetic energy or other
aspects specific to $k$-essence, but simply on the equation of state $w(z)$ and
the sound speed $c_s(z)$ as a function of red shift $z$. The purpose of this
paper is to study possible effects of the sound speed and  its time-variation
on the cosmic microwave background anisotropy and on large scale structure. In
an earlier paper,\cite{Erickson}  the subtle changes due to  sound speed in the
position and shape of the acoustic peaks in the CMB power spectrum  were
investigated and it was shown that the effect of the sound speed could be
discriminated from other physical effects provided that the
\textit{k}-essence\   density is non-negligible (greater than 1\%) at last
scattering.  Here we focus on the large angular scales corresponding to the
quadrupole and first few higher multipole moments.  
We show that $c_s^2 \ll 1$ 
at small red shift can produce a suppression of these first few multipole
moments, an effect similar 
to the low quadrupole 
observed in the COBE
data.\cite{COBE,tegmark} We also show how rapid variations in the sound speed,
as in  the \textit{k}-essence\ example, can produce bumps, oscillations  and
other novel features at  small scales in the mass power spectrum.   The precise
effect depends on the  detailed model.

As noted above, our calculations and examples do not assume that the dark
energy consists of $k$-essence specifically, which is a purely kinetic action
for the scalar field.   In fact, our computer code, a modification of the
standard \CMBFAST\ program, \cite{Sel96,Cal98}  effectively treats the
quintessence component the same as a scalar field with a potential. The code
automatically determines a potential that produces  the desired $w(z)$. 
Independently, the perturbation equations are modified to include the desired
sound speed, $c_s(z)$.  

The effect of the speed of sound on the \CMB\ perturbation equations is such
that for $c_s^2\ll 1$, \kessence\ fluctuations are enhanced via gravitational
instability by the  cold dark matter (\CDM) potentials. For those familiar with
the code, the modifications are straightforward. The perturbed line element is
\begin{equation}
ds^2 = a^2(\tau)[ d\tau^2 - (\delta_{i j} + h_{i j})dx^i dx^j]
\end{equation}
where $\delta_{ij}$ is the unperturbed spatial metric, and $h_{ij}$ is the
metric perturbation. We shall use $h$ to represent the trace of the spatial
metric perturbation. The effect we are examining is due to the perturbations to
the $k$-essence stress-energy in the synchronous gauge for a mode with
wavenumber $k$ (we omit the $k$ subscripts here):
\begin{eqnarray}
\delta\rho &=& -2 \rho \frac{\delta\phi}{\phi}
-(\rho+p)\frac{\delta y }{ y} c_s^{-2}
\cr\cr
\delta p &=& - 2 p \frac{\delta\phi}{ \phi}
-(\rho+p)\frac{\delta y }{ y}
\cr\cr
\theta &=& \frac{1}{\sqrt{2}} k^2 y \delta\phi
\label{flucteqns}
\end{eqnarray}
where $y \equiv 1/\sqrt{X}$ and $\theta$ is the divergence of the fluid
velocity. The density contrast, $\delta \equiv \delta \rho/\rho$, obeys the
equation
\begin{equation}
\dot\delta = -(1+w)\left(\theta + \frac{1}{2} \dot h\right)
- 3 {\cal H}  \left( \frac{\delta p}{ \delta \rho} - w \right) \delta,
\label{deltaeqn}
\end{equation}
where the derivative is with respect to conformal time and ${\cal H} \equiv
\dot{a}/{a}$. The relation between $\delta p$ and $\delta \rho$,
derived using equations (\ref{flucteqns}) above combined with the background
equation of motion, is
\begin{equation}
\delta p = c_s^2 \delta\rho + \frac{\theta \rho }{ k^{2}}
\left[ 3 {\cal H}  (1+w)(c_s^2 - w) + \dot w \right],
\label{deltapeqn}
\end{equation}
which leads to a simplified evolution equation for the velocity gradient
\begin{equation}
\dot\theta = (3 c_s^2 - 1) {\cal H} \theta   + c_s^2 k^2 \delta / (1 + w).
\label{deltateqn}
\end{equation}
We can see in equation (\ref{deltateqn}) that a small sound speed will cause
the velocity gradient to decay; with the conventional gauge choice,
$\theta_{cdm}=0$, the inhomogeneities in the $k$-essence will describe a fluid
which is comoving with the cold  dark matter. From equation (\ref{deltapeqn}),
we see that the second term on the RHS will be negligible even on scales
approaching the horizon. The overall effect is that the pressure fluctuations
$\delta p$ are weak and $k$-essence fluctuations are enhanced via gravitational
instability by the CDM gravitational potentials.
The  \CMBFAST\ code takes $w(a)$ and $c_s(a)$ as inputs, so it is possible to
manually adjust these functions to have any values (including, of course, $c_s
= 1$). 


\begin{figure}
\begin{center}
\epsfig{file=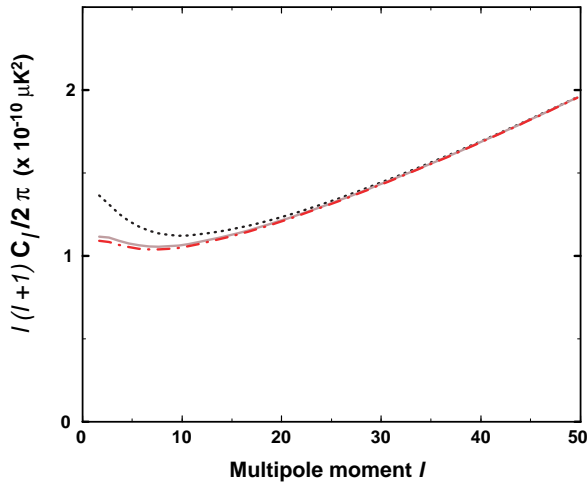,width=3.5in}
\end{center}
\caption{Comparison  of the lowest multipole moments of the CMB temperature
power spectrum for a series of models with $w=-0.8$: (a) $c_s=1$ (dotted); (b)
$c_s=1$ until $z=5$ and then $c_s=0$ for $z<5$ (solid); and (c) $c_s=0$ for all
$z$ (dot-dashed).}
\label{fig:fig2}
\end{figure}

As a first set of cases we consider a sequence of models in which the equation
of state is  constant,  $w=-0.8$, and we vary the sound speed. Note that this
example does not correspond to  tracker-type evolution and, like the
cosmological constant, the  quintessence component becomes negligible at  $z
>2$ or so. Fig.~\ref{fig:fig2} compares three cases: (a) $c_s=1$ (dotted); (b)
$c_s=1$ for $z \ge 5$ and then $c_s=0$ for $z<5$; and (c) $c_s=0$ for  all $z$.
All  models have  $\Omega_m=0.25$, $H_0=70$~km/s/Mpc, and  $\Omega_b= 0.04$,
where $\Omega_{m(b)} $ is the  ratio of the matter (baryon) density to the
critical density and $H_0$ is the Hubble parameter.

We observe the following features:

\noindent
(1) The (late-time) integrated Sachs-Wolfe (ISW) effect is suppressed  if the
sound speed is near zero at recent red shift.  The ISW effect occurs in models
with $\Omega_m<1$ because the gravitational potential decreases as photons
propagate between the last scattering surface and today. The net effect is an
increase in the multipoles  on  angular  scales which enter the horizon when
$\Omega_m<1$, that is to say, the  low-$\ell$ multipole moments.  This is the
effect  predicted for  models with a cosmological constant, for example. Dark
energy with sound speed near zero  suppresses this effect because the
fluctuations  in the quintessence component  add to the gravitational potential
and compensate for the decrease one would obtain if the quintessence were
smoothly distributed.\cite{rdave}

\noindent
(2)  For constant $w$ models the ISW suppression is only weakly dependent on
the time-dependence of the sound-speed.  For example,  the lower two curves in
Fig.~\ref{fig:fig2} nearly overlap.  The reason for this is apparent -- the
quintessence component is a negligible fraction of the energy density for
$z>2$, so the behavior of the sound speed is only important over a narrow range
of recent  red shift.

We have not shown a plot of the multipole moments for large values  of $\ell$
because the curves are essentially degenerate over  this range.  Similarly, the
mass power spectra for these models are virtually identical.  Hence, the only
effect of the sound speed for these non-tracker type models with nearly
constant $w$ is a modest suppression of the low-$\ell$ multipole moments in 
the CMB.

\begin{figure}
\begin{center}
\epsfig{file=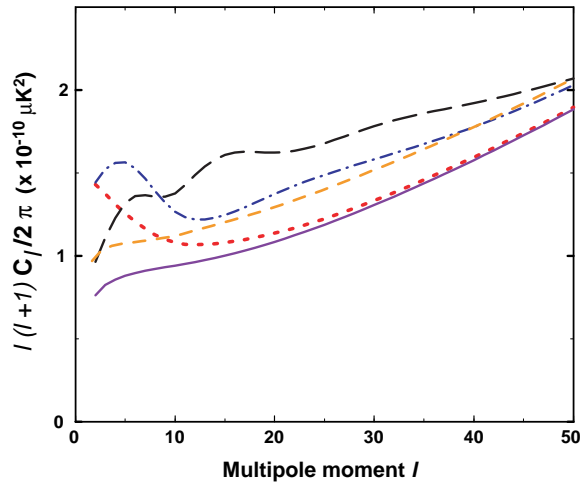,width=3.5in}
\end{center}
\caption{Comparison of the lowest multipole moments of the CMB temperature
power spectrum for a series of models with  the same $w(z)$ (in this case,
corresponding to the $k$-essence model in Fig.~\protect{\ref{fig:fig1}}   and
Ref.\protect{\cite{kess}}) but  different $c_s(z)$: (a) $c_s=1$ (dotted); (b)
$c_s=1$ for $z> 10$ and $c_s=0$ for $z<10$ (solid); the $k$-essence model shown
in Fig.~\protect{\ref{fig:fig1}} (dot-dashed); and two variations (short- and
long-dashed)  with rapid variations in $c_s^2$ (spikes) as described in the
text.}
\label{fig:fig3}
\end{figure}

In Fig.~\ref{fig:fig3}, we show  the small-$\ell$ CMB multipole moments for
models with the same $w(z)$, where $w$ is now  $z$-dependent. As a specific
example, we choose $w(z)$ in these examples to correspond to the $k$-essence
model shown in Fig.~\ref{fig:fig1} and discussed in Ref.\cite{kess}. As
before, all  models have current values of $\Omega_m=0.25$,  $H_0=70$~km/s/Mpc,
and  $\Omega_b= 0.04$.  The details of $w(z)$ are not important here except
that it has the feature that the dark energy is a tracker component that is 
non-negligible at last scattering and earlier, roughly 10\% of the critical
density. We note the following features:

(1) Comparing the models with $c_s=1$ (dotted) and  with the case with  $c_s=1$
for $z>10$ but $c_s=0$ for $z<10$ (solid), there is greater suppression than in
Fig.~\ref{fig:fig2} because the quintessence component  is non-negligible for a
broader range of red shift, enhancing its clustering effect.  (We have
considered this example of time-variation because it is generally plausible 
that  the sound speed of the quintessence component  changes as it overtakes
the matter density, as occurs in the case of $k$-essence.)  

(2)  More general time-variation of the sound speed can produce bumps and
wiggles in the low-$\ell$ multipole moments. The remaining examples in
Fig.~\ref{fig:fig3} are cases where the sound speed  corresponds to the actual
$k$-essence model shown in Fig.~\ref{fig:fig1} (dot-dashed), or simple
variations.   For the variations, we have considered cases where the sound
speed is $c_s=1$ until last scattering, then drops to $c_s=0$ except for a
spike near $z=100$ where the quintessence component approaches the attractor
solution with $w=-1$. While this roughly  mimics  features that are
characteristic  of $k$-essence, here we have bounded  $c_s\le 1$ and treated
the spike as a smoothly varying function to illustrate that the unusual
features seen in the  actual $k$-essence example do not depend on having a
sound speed greater than unity or with  near-discontinuous behavior. We find
that the precise form of the low-$\ell$ power spectrum is sensitive to the red
shift and width of the spike, producing in some cases bumps and wiggles in the
small-$\ell$  multipole moments.

Curiously, the COBE data suggests a suppression, and perhaps even a bump-like 
feature in the low-$\ell$ multipoles.\cite{COBE,tegmark} The data is not
decisive because the cosmic variance and experimental error are large for the
low-$\ell$ multipoles.  Forthcoming data from the MAP and Planck satellite
experiments  may lead to some improvement.

\begin{figure}
\begin{center}
\epsfig{file=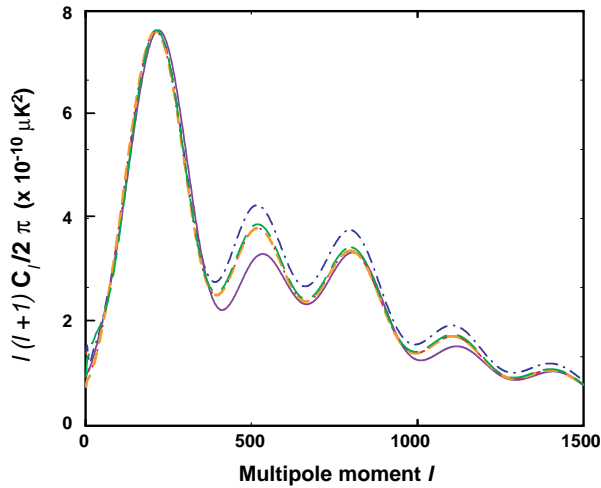,width=3.5in}
\end{center}
\caption{
Comparision of higher multipole moments of the CMB temperature power spectrum
for the models in Fig.~\protect{\ref{fig:fig3}}.  The spectra have been
normalized so that the amplitudes match at the top of the first acoustic peak.}
\label{fig:fig4}
\end{figure}

Fig.~\ref{fig:fig4} shows the full CMB  power spectra for the same set of
models as in Fig.~\ref{fig:fig3}. Recall that these models are completely
identical except  for differences in $c_s^2(z)$.  These  spectra are roughly
similar, but  could be easily distinguished in near future  CMB power spectrum
measurements, {\it e.g.,} by the MAP and Planck satellites. However, detecting
and distinguishing  $c_s^2(z)$ is not as simple as the Fig.~\ref{fig:fig4} may
suggest. The modest differences  in peak heights and positions seen here  can
be approximately mimicked by varying other parameters. So, the challenge
becomes to separate effects of $c_s^2(z)$ from  those of the other
parameters.   As shown in Ref.\cite{kess}, the effects are not completely
degenerate, so highly precise measurements of the power spectra can make the
separation, in principle.  However, the differences are small enough that a
systematic search through possible forms for $c_s^2(z)$ along with the standard
parameters is required in most cases to detect them, and this is not part of
any current fitting procedure.  Observing anomalies in the  low-$\ell$
multipole moments and in the mass power spectrum (see below) as discussed in
this paper would provide strong motivation for developing such a procedure.

We note that similar suppression of the low-$\ell$ multipoles has been obtained
previously by keeping $c_s^2=1$  but varying the equation of state
$w(z)$.\cite{Cal98}  The greater is the value of $w$, the  greater are the
fluctuations in the quintessence  field,  which is the same trend one obtains
by lowering the sound  speed. However, if $c_s^2=1$, the suppression of
multipoles only  occurs if  $w(z)$ is greater than $\sim -0.2$ today. This is 
inconsistent  with the observed acceleration of the universe, which requires
$w(z)< -1/3$.\cite{Bahcall,other}  On the  other hand, keeping $w(z) \approx 
-1$ today and varying the sound speed  history is an interesting alternative 
because it is physically motivated and  does not run into conflict with any
other current observations.

Similarly, one could modify the low-$\ell$ multipoles by altering the 
primordial power spectrum, for example, by introducing an inflation potential
with bumps or dips strategically placed so as to produce  suppression or bumps 
in the low-$\ell$  multipoles.  This solution is unappealing. Introducing such
features at any point  in the inflaton potential  requires very awkward
fine-tuning; but, in this case, where we are seeking to alter only the
low-$\ell$ multipoles, such an approach is even more awkward because the
features in the inflaton potential must be introduced in a way so as to affect
specifically those wavelengths that  recently entered the horizon (rather than 
longer or shorter wavelengths). There is simply no good reason for this.  In
contrast, it is natural to  associate the suppression or bumps in the
low-$\ell$ multipoles with  dark energy because it only dominates at late times
and so will naturally affect modes whose wavelengths  only recently entered the
horizon.

\begin{figure}
\begin{center}
\epsfig{file=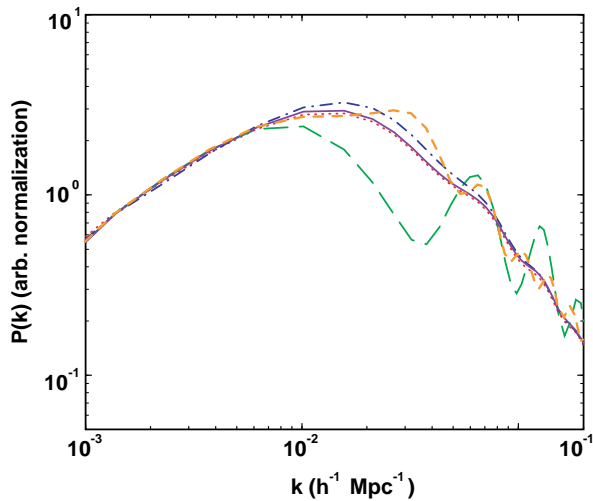,width=3.5in}
\end{center}
\caption{
Comparison  of the {\it shape} of the total fluctuation power spectrum, $P(k)$
as a function of wavenumber $k$ for the sequence of models in
Fig.~\protect{\ref{fig:fig2}}.  The normalization of the curves  is
arbitrary.  
}
\label{fig:fig5}
\end{figure}

Fig.~\ref{fig:fig5} compares the total fluctuation power spectrum
(including contributions from fluctuating components)
for the same models as in Fig.~\ref{fig:fig3}.  We have intentionally included
an example with exaggerated features to illustrate that the sound speed effect
can be quite large (in this case, by making an optimal choice of the red shift
and width of the spike in $c_s$). The main effect to be observed here is that
there can be enhancement in power, suppression on some scales,  oscillations
and other features in $P(k)$ at large $k> 0.01$~h/Mpc. Clearly, uncertainty in
the sound  speed can lead to non-negligible  uncertainty in $\sigma_8$ and,
more generally, the shape of the power spectrum  even when all other parameters
are fixed. The effect could be important in explaining anomalies  that may
arise in $P(k)$ as high quality red shift survey data is obtained and compared
to CMB measurements.


When $c_s^2 =0$, the quintessence fluctuations are enhanced, producing the
effects in the microwave background and mass power spectrum that we have
observed, but the fluctuations do not cluster like matter because the equation
of state, $w$, is different. For example, consider a universe containing only
quintessence with $c_s^2=0$ and perturbations $\delta= \delta\rho/\rho$, where
$\rho$ is the quintessence energy density. Since $w\ne0$, the Jean's
instability equation is modified,
\begin{equation}
a^2 \frac{d^2 \delta}{da^2} + \frac{3}{2} a  A[c_s^2,w] \frac{d \delta}{da}
+\left( {k^2 c_s^2 \over {\cal H}^2} -\frac{3}{2} B[c_s^2,w]\right) \delta =0
\end{equation}
where $a$ is the Friedmann-Robertson-Walker scale factor and $A[c_s^2,w]= 1- 5
w + 2 c_s^2$ and $B[c_s^2,w]= 1 - 6 c_s^2 + 8 w - 3 w^2)$.\cite{Paddy}  
For $c_s^2=w=0$,
we have  $A[c_s^2,w]= B[c_s^2,w]=1$ and we obtain the conventional
dust-dominated  growing solution, $\delta \propto a$.  However, in general, 
the ``growing'' solution is  $\delta \propto a^{\gamma}$ where $\gamma=\frac{1}{2} (1 - \frac{3}{2}A + [ (1-\frac{3}{2} A)^2 + 6 B ]^{1/2})$
, which is less than zero for  $w <-0.12$ (for
all $c_s^2 \ge 0$).  Hence,  despite the fact that $c_s^2=0$, the accelerated
expansion dominates and causes the fluctuations to  decay. In a universe with a
mixture of matter and quintessence,  the matter slows the expansion, so the
quintessence may cluster  {\it when it is subdominant}.   Indeed, it is the
gravitational influence of matter fluctuations on quintessence that is
enhancing the perturbations  in quintessence  that have been studied here.
However, this clustering  is halted once quintessence begins to dominate and the
effective value of $w$ becomes negative.
The only significant effects are the kinds of
changes in the microwave background and mass power spectra seen here.

What we conclude, then, is that modest changes in $c_s(z)$, as motivated by
some existing models of dark energy, can produce modest but measurable changes
in the CMB and mass power spectra. The highly precise data obtained from the
MAP and Planck satellites and from the Sloan Digital Sky Survey may reveal
these subtle effects.  The precise behavior of the sound speed is, by itself, 
of limited interest. But, what is important is that the detection of any 
deviation from $c_s^2=1$ would be a direct sign that the dark energy is a
complex, dynamical fluid rather than an inert cosmological constant. Hence, it
is a target well worth pursuing.

\begin{acknowledgments}
We thank J.P. Ostriker for encouraging this project and 
for many useful conversations. We also thank R. Scherrer for helpful
comments. This work was supported by 
and NSF Graduate Research Fellowship (SD),
NSF grant PHY-0099543 (RRC)
and 
Department of Energy grant 
DE-FG02-91ER40671 (PJS).
\end{acknowledgments}


\begin{thebibliography}{99}

\bibitem{Bahcall} 
For a review, see
N.~Bahcall, J.P.~Ostriker, S.~Perlmutter and P.J.~Steinhardt,
{\it Science} {\bf 284}, 1481 (1999).

\bibitem{Cal98}  
R.R.~Caldwell, R.~Dave and 
P.J.~Steinhardt, Phys. Rev. Lett. {\bf 80}, 1582 (1998);
J.~Frieman, C.~Hill, A.~Stebbins, I.~Waga, Phys. Rev. Lett. 
{\bf 75} 2077 (1995); 
P.J.E.~Peebles and B.~Ratra, Ap. J. Lett. {\bf 325}, L17 (1988);
B.~Ratra and P.J.E.~Peebles, Phys. Rev. D {\bf 37}, 3406 (1988).

\bibitem{kess} 
C.~Armendariz-Picon, V.~Mukhanov, and P.J.~Steinhardt,
Phys. Rev. Lett. {\bf 85}, 4438 (2000); Phys. Rev. D
{\bf 63}, 103510 (2001).

\bibitem{tracker}
I.~Zlatev, L.~Wang and P.J.~Steinhardt, Phys. Rev. Lett. {\bf 82}, 896 (1999); 
Phys. Rev. D {\bf 59}, 123504 (1999).

\bibitem{Erickson} 
J.~Erickson, R.R.~Caldwell, P.J.~Steinhardt, V.~Mukhanov,
and C.~Armendariz-Picon, Phys. Rev. Lett. {\bf 88}, 121301 (2001).

\bibitem{COBE} 
G.~Hinshaw {\it et al.}, Ap. J. Lett. {\bf 464}, L17 (1996).

\bibitem{tegmark}
M.~Tegmark, Ap. J. Lett. {\bf 464}, L35 (1996).

\bibitem{Sel96} 
U.~Seljak and M.~Zaldarriaga, Ap. J. {\bf 469}, 437 (1996);
\texttt{arcturus.mit.edu:/80/matiasz/CMBFAST/cmbfast.html}.

\bibitem{rdave} 
For a more thorough analysis of 
quintessence fluctuation effects, see R.~Dave, U. of Penn.
Ph.D. thesis (2002);
R.~Dave, R.R.~Caldwell, and P.J.~Steinhardt, 
Phys. Rev. D {\bf 66}, 023516 (2002).

\bibitem{other} See, for example,
S.~Perlmutter, M.S.~Turner, M.~White, Phys. Rev. Lett. {\bf 83}, 670 (1999);
R.~Bean, S.H.~Hansen, and A.~Melchiorri, Nucl. Phys. Proc. Suppl. 
{\bf 110}, 167 (2002).

\bibitem{Paddy}
T. Padmanabhan, {\it Structure Formation in the Universe},
Cambridge University Press (1995).




\end{thebibliography}
\end{document}